\magnification=\magstep1  
\vsize=8.5truein
\hsize=6.3truein
\baselineskip=18truept
\parskip=4truept
\vskip 18pt
\def\today{\ifcase\month\or January\or February\or
March\or April\or May\or June\or July\or
August\or September\or October\or November\or
December\fi
\space\number\day, \number\year}
\centerline{\bf Exact Calculations of Membrane Areas with Simple Models.}
\medskip
\centerline{by }
\centerline{J. Stecki }
\vskip 20pt
\centerline{ Department III, Institute of Physical Chemistry,}
\centerline{ Polish Academy of Sciences, }
\centerline{ ul. Kasprzaka 44/52, 01-224 Warszawa, Poland}
\bigskip
\centerline{\today} 
\vskip 60pt

\centerline{\bf{ Abstract}}
The distinction between the true total area and the projected area is 
elucidated with soluble models which represent the membrane 
as a self-avoiding string on a plane. Constraining the total area to a 
predetermined value  changes the averages very significantly. The latter
are calculated exactly from the generating functions of self-avoiding 
walks and are shown as functions of activities $q$ and $r$  related to
temperature $T=\pm 1/\log (q)$ and lateral force $f=-\log (r)$.
The constraint makes the partition functions and averages valid 
for all $q,r >0$ and  reduces the ratio of $A_{tot}$ to the projected
area $L$.  High temperature divergences are supressed. Possible 
applications to simulated bilayers/membranes are discussed.
 
\vfill\eject

{\bf I. Introduction}

 Interfaces and  membranes as well all undergo shape fluctuations due
to thermal motion.
A twodimensional sheet
such as a bilayer or a planar interface 
changes its shape by excursions into the third dimension; thus arises
the distinction between the "true" total area $A_{tru}$ 
and the "projected" area. The latter term has been coined to emphasize  
that the "true" (proper,intrinsic) area of a planar interface 
when projected onto the edge of the volume, produces $A$, 
the projected, or nominal, area.
The definition of the interfacial tension, 
{\it via} the  constant-volume increment $\delta F$ of 
the free energy $F$ 
$$ \delta F =\gamma \delta A~ \eqno(1.1)$$
refers to the projected area $A$ . 
Another expression that both planar interfaces and membranes share,
is the limiting form of the structure factor, dominated by capillary 
waves.

However, there are important distinctions, essentially originating  
from the fact that 
an interface is an open system, also open with
respect  to particle exchange. A liquid surface fluctuates and
varies its surface area  by diffusion of molecules in and out of  
either bulk phase. A surface of a perfect crystal is an extreme example,
as it varies exclusively by particle exchange with vapor phase.

In membranes or bilayers such particle migration does not occur or 
is very rare. 
Commonly membranes and bilayers are formed by self-assembling 
surfactant (am\-phi\-phi\-lic) molecules, embedded in a liquid solvent.
These  are practically insoluble in the solvent forming a variety of micelles 
if not in a bilayer. Consequently the shape 
fluctuations of a membrane or bilayer take 
place {\it under the constraint} of a constant 
particle number $N_s$ forming the surfactant sheet.

 This constraint translates to a constraint of constant membrane area 
 $A_{tru}$. 

 The common formulation of the theory of capillary waves starts with an flat
interface (with its true area $A_{tru}$ originally equal to the 
projected area) which increases its true area by thermal fluctuations. Thus 
the projected area is a given quantity whereas $A_{tru}$ varies in time 
about its thermal average $\langle ...\rangle$  value.

With a membrane one must take the opposite view: 
the true (intrinsic, proper) 
area of a membrane  $A_{tru}$ is a given  quantity whereas the projected 
area is a fluctuating
quantity with a thermal average $\langle ...\rangle$. 

The effects of the constraint of constant particle number in the 
membrane are discussed below.
Introducing these constraints into the theory has proved  difficult;
practically the classical reference$^{1,2}$ is entirely up to 
date$^{3}$ even today. A mention of some difficulties is present in a
theorerical paper$^{4}$. A most useful source of information have been
computer simulations.
  The areas of bilayers have also been extracted from 
simulation data $^{5-7}$. Overall,
the simulations of bilayers have been very numerous$^{5-26}$, but not 
all determined the {\it bilayer isotherms} i.e. lateral tension 
{\it vs.} area dependences$^{5,6,10,12,13,16,21-24}$.

Although the definition (1.1)
 of $\gamma$ may be the same, in fact the behaviour  of $\gamma$ of bilayers
 has been so different from $\gamma$ of interfaces/surfaces - that a
 specific terminology of "lateral tension" was introduced.
The "exotic" properties of the lateral tension 
have been recently enumerated$^{26}$.

In this paper we find useful insights from exact solutions of simple models.
The membrane is modelled as a onedimensional string embedded in a square
lattice. Fig.1 shows such a string and the two areas associated with it. 
As calculations show, besides the  the membrane/bilayer case, of chief 
interest,
the interfacial situation can also be modelled with appropriate choice 
of parameters while removing the constraint.

 In Section II we apply the known generating functions$^{27}$ to derive some 
new variants of these and then to calculate the averages: the average
projected area for given full area (the membrane) and the average total
area for given projected area (an interface); we illustrate the 
spectacular differences with Figures. Finally we consider the case of 
a boxed membrane/bilayer, such as is modelled in molecular dynamics 
simulations, for which both areas, total and projected, are given and
fixed. Section III is the summary and a discussion, illuminating {\it i.a.} 
the issue of the bending (rigidity) coefficient.  
\vfill\eject

{\bf II. The Random Walk and its Generating Functions.}

We model the membrane as a twodimensional string in an infinite square
lattice. The string is generated by a partially directed 
self-avoiding random walk (PDSAW). With lattice constant $a_0=1$, 
the allowed steps are: up, down, or to the right, {\it i.e.}   $(0,+1)$, 
$(0,-1)$, or $(+1,0)$, respectively. The total number of steps,  $N$, is made of 
$L$ horizontal $x-$steps, and of $A_v$ vertical steps.  $A_v=n_+ + n_-$ is
made of positive and negative steps $n_+\ge 0$ and $n_-\ge 0$.
Thus always $N=A_v + L = n_+ + n_- + L$. Such a string is shown in Fig.1.

A string of length of $N$ steps extends from (0,0) i.e. from $x=0, y=h_1=0$,
to $(L,h_{last})$ i.e.to $x=L,y=h_L=h_{last}$. 
If the string is to represent a onedimensional membrane,
the length $N$ becomes the membrane area, and $L$  the {\it projected
area}. The usual definition of an area of a smooth surface translates
into $ s^2 =\Delta x^2 + \Delta y^2$
 as the surface element produced by one step;
 we approximate the hypothenuse $ s$
by the sum $s=\Delta x+\vert \Delta y\vert$.
Thus the area is always somewhat overestimated, but  
the difference with the usual definition was found to be quite small.
Then the total area ("true") is $N$, $A_{tru}\equiv N$, and $N=A_v+L$.  
In the membrane picture $N$ once given is constant; it does not fluctuate.
In the interface model, $L$ is given and constant and $N$ can take any value
$N\ge L$.

 The generating function is constructed as
$$ G = \sum_{all~walks} q_-^{n_-}\cdot q_+^{n_+}\cdot r^L  \eqno (2.1)$$
where  $r, q_+,q_- $ are weights assigned to these steps and counting 
parameters. Generally up (plus) steps are assigned the same weight as the
down (minus) steps, i.e. $q = q_- = q_+$ and $q^{A_v}$ is substituted for
$q_-^{n_-}\cdot q_+^{n_+}$.

The parameter
$r>1$ will favor large $L$, i.e. extended, relatively flat configurations,
whereas $r<1$ will favor small $L$. The parameter $q<1$ will
favor less deviations from flatness, whereas $q>1$ will favor configurations
with many folds  - which are suggestive 
of a crumpled membrane or floppy bilayer. Restriction of $h_{last}$ to 
the value $h_{last} = h_1 = 0$ may also be imposed in order to have 
periodic boundary condition in  the $x$-direction.

A still more general model would introduce the bending parameter $w$ and a
factor $w^T$ where $T$ is the number of turns. Here we take $w=1$, i.e. we
ignore the number of turns. The issue of bending coefficient in the 
context of this model is discussed in Section III. 

  Clearly (2.1) can be interpreted as a grand partition function with
$r, q$ as activities; thus 
$$ q\equiv \exp [-\beta\epsilon_v] \eqno(2.2)$$ 
where $\beta\equiv 1/kT$ and $\epsilon_v$ is the energy cost of a vertical step.
Now we note that $q<1$ implies $\epsilon_v>0$ i.e. the choice for 
{\it modelling an interface} which deviates from flatness at some cost. This 
cost is reflected in the positivity of $\gamma$ defined by (1.1). 
Otherwise, $q>1$ and $\epsilon_v<0$ owing to attractive intermolecular forces 
between the segments of the bilayer or membrane. 

The activity $r$ may be interpreted as $r\equiv \exp[-\beta\epsilon_h]$ 
with some horizontal energy cost, but if we reintroduce the lattice 
constant $a_0$, then we see that $r^L$ may be profitably interpreted as 
$$ r^L =\exp[-\beta f a_0 L] \eqno(2.3) $$
where now $L$ is a nondimensional integer and $f$ is a {\it force} in the 
$x-$direction. Then $r>1$ favoring large $L$ implies a pulling force
(which is negative) whereas $r<1$ implies a compressing force 
(which is positive);  $r=1$ is the locus of the {\it tensionless states}.
 
We adapt to our purposes the general solution$^{27}$ for $G$ 
$$G=(r +q_- +q_+ -(r+2)q_-q_+)/(1 - r - q_+ - q_- + q_-q_+(1 + r))\eqno(2.4) $$
(with $w=1$ as mentioned above).
Without distinction between positive and negative vertical steps,
i.e. with  $q_-=q_+=q$, 
$$ G_o = (2q + r + q r)/(1 - q - r - qr) \eqno (2.5)$$
With the condition of at least one horizontal step, $G$ is$^{27}$
$$ G_1 =   ( r + q r)/(1 - q - r - q r) \eqno (2.6)$$
 All  three generating functions impose no restrictions on the final position
$(L,h_{last})$ attained after all $N$ steps; 
$$h_{last} = n_+ - n_- + h_1 = n_+ - n_- ~.\eqno (2.7)$$
For strings with periodic boundary condition, taken with
Fourier analysis in view, the walk is from  $(0,0)$ to 
$(L,0)$, $h_{last}=0$ . The appropriate generating functions with this
constraint embedded, are derived below.

The generating function $G_o$ counts all horizontal $L$ steps and 
all vertical $A_v$ steps which produce configurations of the string; 
the latter can be gathered together in the 
combinatorial factor $g(L,A_v)$
$$ G_o(q,r) = \sum_L\sum_{A_v} g_o(L,A_v) q^{A_v}r^L~~~~N=A_v+L \eqno (2.8)$$
and similarly  $G_1$.

{\bf  Canonical averages for an interface}.

The shape fluctuations of an interface take place at
$L$ fixed and constant.  The statistical averages, e.g. 
$\langle N\rangle =\langle A_v\rangle +L$
are calculated under this condition. From the  generating
functions, the Taylor expansion of the generating function
$$ G_o=\sum_L Z(L,q) r^L  \eqno (2.9)$$
produces the partition function $Z(L,q)$.  The averages are found from  
$$ \langle A_v\rangle = q(d\log Z(L,q)/dq)  \eqno (2.10)$$
and $\langle N\rangle  = L + \langle A_v\rangle$.
The vertical part of $N$, $A_v$, can be any 
nonnegative integer. In this way, for $L\geq 2$,  from (2.5-2.6) 
$$ Z(L,q)=((1+q)/(1-q))^{(L+1)}.   \eqno(2.11)$$
The same $Z$ obtains from $G_1$. The average total area $N$ follows
from (2.10)
$$\langle N \rangle = L+\langle A_v \rangle = L+2qL/(1-q^2). \eqno(2.12)$$

 The  areas are best represented by a normalized quantity 
${\cal L} = L/N = L/(A_v + L)$; always ${\cal L}\in [0,1]$. 
Rewriting (2.12)
$${\cal L} = L/(L+\langle A_v\rangle )=f_1(q) =(1-q^2)/(1+2q -q^2).\eqno(2.13)$$
This expression results from both $G_1$ and $G_o$.
Since $L$ is constant, 
${\cal L}$ is $ L/\langle N\rangle = L/(L+\langle A_v\rangle)$.

For the canonical averages at given $q,r$, i.e. without fixing $L$ (or $N$) 
at a predetermined value, $G_1$ gives $f_1(q,r)$, (2.13) again; 
However, from $G_o$ by using 
$$ \langle L \rangle = r(dG_o/dr) / G_o \eqno(2.14)$$
one obtains
$${\cal L}_o = o_1(q,r)=(1 + q)^2r/(2q + r + 2qr + q^2r). \eqno(2.15)$$
This  does not differ much from (2.13) and is quoted only for completeness.
The independent variables were $L,q$ or $q,r$; the length $N$ and $A_v$ 
resulted as averages, so that $\cal L$ was
 $\langle L\rangle /(\langle L\rangle +\langle A_v\rangle $. 

Keeping the distinction between $q_+$ and $q_-$  and expanding (2.1) 
in powers of $r$, we find $Z(L,q_+,q_-) = \Psi^L$ where 
$$\Psi = { {1-q^2} \over {1 - q_+ - q_- +q^2}}. \eqno(2.16) $$ 
and the expansion of $Z$  is
$$Z(L,q_+,q_-) = \sum g(n_+,n_-,L) q_+^{n_+} q_-^{n_-}\eqno(2.17) $$
The walk beginning at (0,0) and ending at  $(L,h_{last})$ will have
$h_{last}=h_{last}-h_1 = n_+ - n_-$.
Let us call this quantity $n$. Then in the sum we insert a Kronecker Delta
ensuring that $n = n_+ - n_-$; by using its Fourier representation,
putting $q_+=q_-=q$, we obtain
$$Z(n;L,q) = (1/\pi ) \int _0^\pi dk\cos (n*k)({{1-q^2}\over 
{1-2q\cos (k)+q^2}})^m \eqno(2.18)$$
where the power $m$ is $L,L+1,L-1$ depending on the starting generating
function. The result of integration depends on the inequality 
$(1+q^2) > \vert 2q\vert$ which is fulfilled for all $q>0$ except for
one point $q=1$. This point approached 
from below corresponds to the infinite temperature in the interface models 
with $q=\exp(-\beta\epsilon_v)$,
$\epsilon_v>0$;  approached from above corresponds to the infinite temperature 
in membrane models for which $\epsilon_v<0$. 
For $n=0$ in (2.18) we obtain the partition function for an interface with 
periodic boundary condition; it reads
$$Z(0;L,q) = P_{m-1}(u) ~~~ u\equiv (1+q^2)/(\sqrt{1-2q^2+q^4})  \eqno(2.19) $$
Here $P_m(u)$ is the Legendre polynomial, $u \ge 1$. The normalized 
average area $\cal L$ 
is in this case $L/(L+\langle A_v\rangle)$ and $\langle A_v\rangle$ is 
calculated after (2.10) as 
$q(d\log Z/dq)$; we find then that we must take $q<1$. Finally
$${\cal L} =  (1+u - P_{L-1}(u)/P_L(u) )^{-1}. \eqno(2.20)$$
As we shall see below, this quantity behaves quite unlike those $\cal L$'s 
constrained by the total length $N=L+A_v$.
It agrees very closely with $f_1(q,r)$.

\vfill\eject

{\bf Canonical averages for a membrane/bilayer.}

Alternatively, treating 
the string as a membrane, we impose a fixed length $N$.
This constraint is introduced into $G$'s.
The new generating function contains a selection of terms from (2.5),
such that $L+A_v=N=const.$,
$$ G(N;q,r) =\sum_{L=1}^N g(A_v=N-L,L) q^{N-L}r^L~~~N=A_v+L=const. \eqno(2.21)$$
There are two cases possible: either all $L$ compatible with given $N$
are allowed, or both $L$ and $N$ are
given prescribed values. In the latter case the expansion (2.21)
reduces to one term, with $(A_v=N-L,L)$. 
From the expansion (2.8) of $G_o$, in powers of $r$ and then $q$, 
we find $g_o(A_v,L)$;
alleviating the notation by writing $a\equiv A_v $,
we write the explicit result 
$$ g_o(a,L)=\sum_{m=0}^a {L\choose m}{a-m+L-1 \choose L-1}~~~  \eqno (2.22)$$
For $a>L$ the upper limit of the sum is $L$.
With $g_o$ known, we can construct the partition function $Z_o$ as a polynomial 
in $q$ and $r$,
$$Z_o(N;q,r)=\sum_{L=1}^N g_o(N-L,N)\times q^{(N-L)}\times r^L  \eqno(2.23)$$
by explicit substitution of (2.22) into (2.23).
Now $\langle L\rangle $ and $\langle a\rangle $ follow explicitely by
(2.10) and (2.14).
The same procedure works for $G_1$.
For these averages at $q,r;N$ always ${\cal L} 
= \langle L\rangle/(\langle L\rangle +\langle Av\rangle) $.

A direct and neat way is to introduce a new counting  parameter $t$ for every 
step whatever it is and expand 
$$ G_o(t;q,r) = (2qt + rt + q rt^2)/(1 - qt - rt - qrt^2) \eqno (2.24)$$
in powers of $t$, 
 $$ G_o(t;q,r) = \sum_{t=1}^\infty Z_N(q,r) t^N~~~~ (N=A_v+L). \eqno(2.25) $$
$Z_N$ is used for calculations of average $\langle L\rangle$ (which must be
equal to $N-\langle A_v \rangle$.

Finally, pinning both ends of the string in order to have a 
possiblility of p.b.c., is done in two alternative ways:
either we expand (2.4) in powers of $t$,  select the coefficient of $t^N$
which is $Z_N(q_+,q_-,r)$; substitute there $q_-\rightarrow q/e$ and 
$q_+\rightarrow qe$, expand in powers of $q$ and $e$ and select
the coefficient of $e^0$ - it is the partition function $Z_N(q,r)$ for 
both ends pinned, useful for p.b.c.. Alternatively, the equality of 
$n_+$ and $n_-$ is enforced via a Kronecker Delta and its Fourier
representation as an integral.  
 Either way, its generating function is found with a rather unusual appearance 
$$G_e(q,r) = r*\sqrt{B};  B\equiv (1-q^2)/((1-q-r-qr)(1+q-r-qr)).\eqno(2.26)$$
To calculate the averages at constant $N$, the same procedure is used as for $G_o$. Put
$q\rightarrow qt, r\rightarrow rt$, extract the Taylor series coefficient
of $t^N$ and use (2.10),(2.14). The resulting $le(q,r)$ depends on $N$.
  
These exact enumerations 
can be pushed very far to $N\gg 100$ - but then simple 
asymptotic expressions take over. These are derived now.

Both $G_1$ and $G_o$ are positive and finite for small $q$ and $r$; their
divergences are determined by zeros  of the denominators. This 
condition, $1 - r - q - q r = 0$, determines a line in the $q,r$ plane 
cutting diagonally from $(0,1)$ to $(1,0)$. 
 We write the denominator as
$\alpha(t-t_1)(t-t_2)$ where $t_{1,2}$ are its roots and split $G_1$ as
$$G_1 = B_1/(t-t_1) +B_2/(t-t_2) \eqno(2.27)$$
The root $t_1$ is always positive and $t_1<1/r$; the other root is 
always negative. $G$ takes largest values when $t$ is close to $t_1$;
expanding in powers of $t$ (as $(1-t/t_1)^{(-1)} = 1+(t/t_1)+\cdots $)
we see that
the root $t_1 > 0$ is the one that matters. From this new
form of $G_1$ we select the coefficient of $t^N$ and calculate 
$\langle L\rangle $, taking the asymptotic limit of large power of $t$,
i.e. large $N$. Finally we obtain
$${\cal L} \equiv f_2(q,r) = (1/2) +(1/2)(r-q)/\sqrt{(r^2+6rq+q^2)}
 ~~~N=const.\rightarrow\infty.   \eqno (2.28)$$
This expression, like (2.22-2.24), 
is valid in the entire quarter-plane $(q\ge 0, r\ge 0)$. 
$\cal L$ from $G_o$ attains the same asymptotic result in the limit
 $N\rightarrow\infty$. 

An important point is that the partition functions
for fixed $N$ (total area, or length of the string) $Z_N(r,q)$ are positive and
well-behaved for all $q>0,r>0$ whereas the generating functions $G(r,q)$, 
the partition functions for fixed $L$, $Z(L,q)$ and the averages and 
fluctuations derived therefrom, are well behaved and physically acceptable 
only in certain regions of the quarter-plane.

 Fig.2  shows five functions $\cal L$:
(1) the canonical average at given $q,r,N$, denoted as $ll(q,r)$; 
eq.(2.25),(2.10),(2.14);
(2) as (1) but under the periodic boundary condition (i.e.pinning both
ends at $h_{last}=h_1$) - function $le(q,r)$, eq.(2.26)ff;
(3) the asymptotic limit of (1) for $N \rightarrow\infty$, denoted as
$f_2(q,r) $, eq.(2.27).
(4) the canonical average at fixed $L$  (and given $q$),  any $N$, 
any $h_{last}$; function $f_1(q,r)$, eq.(2.13).
(5) as (4) but under the periodic boundary condition $h_{last}=h_1$; 
eq.(2.20).

In order to be physically acceptable, each  $\cal L$ must be 
${\cal L} \in [0,1]$
and also the averages $\langle L\rangle$ and $\langle A_v\rangle$ must
be positive. 
This selection of 5 functions ${\cal L}$ are shown plotted against $q$, for 
a series of values or $r$. The regions $q<1$ and  $q>1$ are smoothly
joined by all averages calculated at constant $N$. The function $f_1$
behaves quite differently: its physical branch is limited to 
the region $0\le q\le 1$.
$f_1$ does represent an interface; the other three - a membrane. 
Similarly under the 
periodic boundary condition ${\cal L}(q,L)$, (eq.2.20),
for either $L=10$ or $L=140$ does not differ visibly (on the scale of the 
plot) from $f_1$; it also represents an interface.
All functions start from $A_v=0$ at $q=0$ but the interface fluctuations
give unbounded $\langle A_v\rangle$ at $q\rightarrow 1^-$ whereas the membrane
fluctuations are bounded.
 The three constant-$N$ functions $ll(q,r;N), le(q,r;N)$ and $f_2(q,r)$ 
cross smoothly the point $q=1$; all three are not too far away from each other. 
In particular the asymptotic $f_2$ would be merged with the other two on the 
scale of the Figures if the plots were not drawn for a choice of small $N$.
 
 The interval  $0\le q <\infty $ is split in two and in Fig.3 amd 4
the plots are against inverse temperature, against $b= -log(q)$ for $q<1$; 
against $b=\log(q)$ for $q>1$, respectively.
In Fig.3 the limiting values  for $b \to 0$ (the limit of high temperatures)
depend on the parameter $r$,
except for the two cases of fixed projected area $L$ where the limit is
${\cal L} =0$, as $L$=const. and $\langle A_v\rangle$ is unbounded. 
The point $b=0^+$ corresponds to $q\to 1^-$.   $f_2(q=1,r)$ 
(eq.2.27) is continuous and  $f_2(1,1)=1/2$. All curves, fixed $N$ or fixed $L$, merge
towards ${\cal L}=1$ as $b$ increases, i.e. $kT\to 0$. 
Obviously, $\langle A_v\rangle$ vanishes there.
In Fig.4 the function $f_1$ is shown and how it takes unphysical values either
negative or larger than 1. The other $L-$fix average(eq.(2.20) is also 
unphysical in this region, unlike the three constant-$N$ functions.  
At very low temperatures all three $\cal L$'s tend to zero because
$A_v$ dominates; $q\gg 1$ favors configurations with many folds for 
which $A_v \gg L$.
 
We have chosen to convert all averages to $\cal L$'s for convenience but 
in this way some dramatic differences in behaviour are hidden from view.
Thus ${\cal L}=0$ implies unbounded $A_v$ at finite $L$ or vanishing 
$\langle L\rangle$ at finite $A_v$; ${\cal L}=1$ if $A_v=0$. In the limit of high
temperatures, the interfacial averages, i.e. those taken at constant $L$,
diverge, whereas the membrane averages i.e. those at constant $N$, produce
finite values (depending on $r$).

Next we change the role of variables and plot below in Fig.5
the same functions against
$r$ at several constant values of $q$; the logarithic scale implies that we
choose  force $f=-\log(r)$ as the independent variable (the ordinate). 

{\bf A boxed  membrane.}

In simulations most often not only the number of molecules forming the
bi\-lay\-er\-/\-mem\-bra\-ne is kept constant, 
but also the membrane or bilayer is confined 
to the simulation box of dimensions $L_x \times L_y \times L_z$.
For our model this implies that not only $N$ is given and constant but
also  $L$. Because in this particular model 
the sum $A_v + L$ coincides with $N$, at any given temperature
 the partition function and the free energy are
$$ Z = g(A_v,L)q^{A_v}~~~   (A_v=N-L, N=const., L=const.) \eqno(2.29)$$
$$ \beta F = -\log g(A_v,L) -A_v\log q \eqno(2.30)$$
The activity $q\equiv \exp[\epsilon/kT] > 1$.
For comparison, a boxed interface  in an identical simulation box,
does not keep $A_v+L$ constant and therefore
$$ Z_{int} = \sum_a g(a,L)q^a ~~~  (L=const.) \eqno(2.31)$$
$$ \beta F_{int} = -\log Z_{int}   \eqno(2.32)$$
The lateral tension of the membrane results as
$$ \hat\Gamma =({\partial F \over \partial L})_{T,N} = -(1/g)(dg/dL)_N \eqno(2.33)$$
The derivative $dg/dL$ of $g(A_v,L)$ is taken at constant $N$ .
Fig.5 shows $\cal L$ for a series of $q$'s with constant $N$ plotted 
against the negative of the force $(-f)=\log (r)$. The asymptotic 
$f_2(q,r)$ (for $N \to \infty$) is also included - it has very much the 
same shape, only shifted along the ordinate according to the value of $q$,
just like all the finite$-N$ curves. For comparison also $\hat\Gamma$ 
from finite differences $(\Delta g/\Delta L)_N$ is shown demonstrating 
the equivalence of ensembles. The continuity w.r.to $q$ is also seen. 
The pulling force raises $\langle L\rangle$ as expected and conversely; the effect 
of $q$ is to shift the entire
curve $L(f)$ to the right or to the left, without much of a visible change of shape.
This can be understood by looking at the maximum term in the canonical 
sum for which the condition is $ (d\log g/dL)_N = -f -\log q$. The parameter
$q$ shifts the curve $f^*(L)$ by a constant amount. 

{\bf III. Summary and Discussion. }

The  model allows explicit calculations in terms of simple 
algebra. Its use  made it possible to demonstrate  the role of the 
membrane-constraint in the statistical mechanics of membranes or/and bilayers.
The constraint of constant particle number in the membrane - no particle
loss or gain to or from the surrounding solvent - calls for a corresponding
approach in the calculation of the membrane thermal fluctuations. In this 
model this translates into a constant length of the string. Such a constraint
is most natural for modelling of a polymer and we expect that the 
calculations for a twodimensional membrane embedded in three dimensions,
will be in effect calculations for a two-dimensional polymer.

 In the course of calculations we have found that relaxing the constraint 
 of constant $N$ and replacing it by a constraint of a fixed projection
 area $L$, leads to a very different behaviour of the averages and -
 a posteriori obviously - modelling the fluctuations of an interface.
 Thus in the same model we have the region $q>1$ inaccessible to interfaces 
 but typical for a membrane - and the region $q<1$ typical for interfaces
 and also accessible for a membrane with a changed sign of the energy 
 of interaction. 

The equivalence of ensembles operates fully in the limit of infinite 
systems but not for finite sizes.  

The force, pulling or compressing the string, has a direct analogy in
three dimensions, as the lateral tension of the bilayer. Cutting a 
suitably chosen part of an S-shaped curve from Fig.5, one can get a
picture very similar to the "bilayer isotherm" (which is a plot of 
lateral tension against the projected area). Curiously, a non-symmetric
derivative $(dg/dL)_N$ i.e. $(g(L_0+1)-g(L_0))/g(L_0)$ assigned to
the point $L_0$, produces such curves.

Because the string is confined to the square lattice, it has an implicit 
resistance to bending which is also seen in the play of the parameter $w$.
The latter has been introduced in the generating functions of the PDSAW$^{27}$ 
in the form of $w^T$ where $T$ is the number of turns; see also$^{28-30}$.
Clearly $w \gg 1$ will 
favor the occurence of multifolded configurations with many turns - very 
low bending coefficient, low resistance to bending. Conversely, $w\to 0$
will favor "no turns" i.e. flat membranes which are obtained with 
large bending coefficients -
large resistance to bending. Thus it follows that our choice of $w=1$ 
does not mean "zero bending coefficient" - and the explanation lies in the 
implicit resistance to bending imposed by the lattice. The 
notions of "semiflexible" and  "super-flexible" strings have been 
introduced$^{28-30}$.

 In this context  it becones necessary to mention the existence of a
sizable  amount of very successful work on exact enumeration - for models 
of polymers as strings confined to a lattice$^{28-34}$. References to 
earlier work can be found there as well. However, 
the issues tackled in 
this paper were not touched in those References, where the emphasis was 
entirely on the important topics of phase transitions, singularities, 
and critical scalings.    
 
Although the usual picture of a membrane assumes an attractive interaction
between its segments - which leads to $q>1$, it is possible to envisage
a string held together by other (intramolecular) forces while the segments
repel each other or a good solvent encourages extended configurations;
this leads to $q<1$ - still a membrane if $N$ is held constant.  

Incidentally we also confirm the view$^{26}$ of the tensionless state $f=0, r=1$
as just another point on the $r$ axis or line on the $q,r$ plane 
without any special features. Contrariwise, $q=1$ is a very special 
point as discussed in Section II and shown in the Figures.
\vfill\eject

{\bf References.}

\item {$^{1}$}  W. Helfrich and R. M. Servuss, Nuovo Cimento 3D, 137 (1984);  
\item {$^{2}$}  W. Helfrich, in {\it  Les Houches, Session XLVIII, 1988,
           Liquids at Interfaces} (Elsevier, New York, 1989).
\item {$^{3}$}   A. Adjari, J.-B. Fournier, and L. Peliti,  Phys. Rev. Lett. 
          {\bf 86}, 4970 (2001). 
\item {$^{4}$}  H. A. Pinnow and W. Helfrich, Eur.J. Phys. E{\bf 3}, 149 (2000);
    for a related conclusion cf. Y. Nishiyama, Phys. Rev. E {\bf 66}, 061907 (2002).
\item {$^{5}$}  W. K. den Otter, J. Chem. Phys. {\bf 123}, 214906 (2005).
\item {$^{6}$}  Hiroshi Noguchi and Gerhard Gompper, Phys. Rev. E {\bf 73}, 021903 (2006),
 where references to earlier work on triangulations  can be found. 
\item {$^{7}$}  A. Imparato, J. Chem. Phys. {\bf 124}, 154714 (2006).
\item {$^{8}$}  B. Smit, Phys. Rev. A {\bf 37}, 3431 (1988).
\item {$^{9}$}  B. Smit, P.A.J. Hilbers, K. Esselink, L.A.M. Rupert,
N.M. van Os, and A.G. Schlijper, J. Phys. Chem. {\bf 95}, 6361 (1991).
\item {$^{10}$} R. Goetz and R. Lipowsky, J. Chem. Phys. {\bf 108}, 7397 (1998).
\item {$^{11}$} G. Gompper, R. Goetz, and R. Lipowsky, Phys. Rev. Lett. {\bf 82}, 221 (1999).
\item {$^{12}$} A. Imparato, J. C. Shilcock, and R. Lipowsky, Eur. Phys. J. E{\bf 11}, 21 (2003).
\item {$^{13}$} A. Imparato, J. C. Shilcock, and R. Lipowsky, Europhys. Lett. {\bf 69}, 650 (2005).
\item {$^{14}$} O. Farago, J. Chem. Phys. {\bf 119}, 596 (2003); for further work on this
      special model cf. O. Farago and P.Pincus, ibid. {\bf 120}, 2934 (2004).
\item {$^{15}$}  G. Ayton, S. G. Bardenhagen, P. Mc-Murty, D. Sulsky, 
and G. A. Voth, J. Chem. Phys. {\bf 114}, 6913 (2001).
\item {$^{16}$}  S. E. Feller and R. W. Pastor, J. Chem. Phys. {\bf 111}, 1281(1999).
\item {$^{17}$}  S. J. Marrink and A. E. Mark, J. Phys. Chem. {\bf 105}, 6122 (2001).
\item {$^{18}$}  E. Lindahl and O. Edholm, Biophysical Journal {\bf 79}, 426 (2000).
\item {$^{19}$}  W. den Otter and W. Briels, J. Chem. Phys. {\bf 118}, 4712 (2003).
\item {$^{20}$}  J. Stecki, Intl. J. Thermophysics {\bf 22}, 175 (2001).
\item {$^{21}$}  J. Stecki, J. Chem. Phys. {\bf 120}, 3508 (2004).
\item {$^{22}$}  J. Stecki, J. Chem. Phys. Comm. {\bf 122}, 111102 (2005).
\item {$^{23}$}  J. Stecki, J. Chem. Phys. {\bf 125}, 154902 (2006).
\item {$^{24}$}  I. R. Cooke and M. J. Deserno, J. Chem. Phys. {\bf 123}, 224710(2005); 
                 (see also http://arxiv.org/cond-mat/0509218).
\item {$^{25}$}  G. Brannigan and F. L. H. Brown, J. Chem. Phys. {\bf 120}, 1059 (2004).
\item {$^{26}$}  J. Stecki, J. Phys. Chem. B 2008, 112(14), 4246-4252.
\item {$^{27}$}  V. Privman and N. M. Svrakic, {\it "Directed Models of Polymers, 
       Interfaces, and Clusters: Scaling and Finite-Size Properties"}, 
        vol. 338 of Lecture Notes in Physics, Springer Verlag, Berlin, 1989. ( esp. pp.15ff).
\item {$^{28}$} Haijun Zhou, Jie Zhou, Zhong-Can Ou-Yang, and Sanjay Kumar,
                Phys. Rev. Lett. {\bf 97}, 158302 (2006).
\item {$^{29}$} Sanjay Kumar, Iwan Jensen, Jesper L. Jacobsen, and Anthony
                J. Guttmann,   Phys. Rev. Lett. {\bf 98}, 128101 (2007). See
                also the preprint no.0711.3482v1 made avaliable at 
                http://arXiv.org/cond-mat. Numerous references to earlier
                work of these authors can be found there.
\item {$^{30}$} A. L. Owczarek and T. Prellberg, preprint no. 0709.3178;
                made avaliable at http://arXiv.org/cond-mat. 
                References to  earlier work can be found there.
\item {$^{31}$} A. L. Owczarek and T. Prellberg,  Phys. Rev. E {\bf 67}, 032801 (2003).
\item {$^{32}$}  T. Prellberg, J. Phys. A {\bf 28}, 1289 (1995)
\item {$^{33}$}  R. Brak, A. L. Owczarek, and T. Prellberg, J. Stat. Phys.
                 {\bf 76}, 1101 (1995); ibid. {\bf 72},737(1993).
\item {$^{34}$} S. Kumar and D. Giri   Phys. Rev. E {\bf 72}, 052901 (2005).

\vfill\eject

{\bf  Figure Captions.}
 
Caption to Fig.1 

The string produced by a random walk on the square lattice. The $x$ coordinate
measures the projected distance-area $L$, the y-coordinate is called "height"
in the text. Here the walk starts at $h_1=0$ and ends at $h_{last}=h_1$.
In the Figure the number of "horizontal" steps is 14, 
$n_+=n_-=A_v/2= $, and the walk is a PDSAW. 
 
Caption to Fig.2

Normalized ${\cal L}$,  $\langle L\rangle/N$, 
$\langle L\rangle/\langle N\rangle$ 
or $L/\langle N\rangle$, as function of $q,r$,
plotted against $q\in [0,+\infty]$ for several values of $r$.
Thick solid line: function $f_1$ eq.(2.13) - unrestricted
canonical average at given $(q,r)$ coincidentally equal to
average at fixed $L$. Diamonds: fixed $L=140$ and pinned at 
$h_1=h_{last}=0$ ( for the p.b.c.);eq.(2.20).
Membrane averages for fixed $N$: thin broken lines - asymptotic
function $f_2(q,r)$ eq.(2.27); thick broken
lines - function $le(r,q)$, eq.(2.26) and ff. text; thin lines - 
function $ll(q,r)$ -canonical average under the constraint of 
imposed value of $N$; here a small value of $N$=10 is chosen to 
emphasize the differences.
All three $\cal L$'s are moving smoothly
with change of the parameter $r$; the Figure shows $r=0.1,1.,9., 30.$
from lowest values to highest. See text.
 
Caption to Fig.3

Normalized $\langle L\rangle/N$ or $\langle L\rangle/\langle N\rangle$ 
or $L/\langle N\rangle$ plotted against
inverse temperature $b=-\log q = \epsilon /kT$ for $0<q<1$.
See Caption to Fig.2.Thick solid line: function $f_1$ eq.(2.13), fixed L;
Diamonds: fixed $L=140$ and p.b.c. ($h_{last}=0$), eq.(2.20); the 
differences with $f_1$ cannot be seen on the scale of this graph.
Thin broken lines - asymptotic function $f_2(q,r)$ eq.(2.27); thick broken
lines - function $le(r,q)$, eq.(2.26) and ff. text; thin lines -
function $ll(q,r)$ both  for a small value of $N$=10. The high-temperature
limits depend on $r,q$, except for fixed $L$ averages where the limit 
is zero. The low temperature limit is unity for all  five $\cal L$'s.

Caption to Fig.4

Normalized $\langle L\rangle/N$ or $\langle L\rangle/\langle N\rangle$ or $L/\langle N\rangle$ plotted against
inverse temperature $b=+\log q = \epsilon /kT$ for $1<q<\infty $.
see Caption to Fig.2,3. Thick solid line: function $f_1$ eq.(2.13), fixed
L ; all values unphysical. The other $\cal L$ at constant $N$:
with thin broken lines - asymptotic function $f_2(q,r)$ eq.(2.27); thick broken
lines - function $le(r,q)$ for pinned string, eq.(2.26) and ff. text; thin lines -
function $ll(q,r)$ for floating string - both for a small value of $N$=10. The high-temperature
limits vary with $r,q$. ?The low temperature limit is unity for all three
averages under constant $N$. ?The low temperature limit is
zero, as $q\gg 1$ favors multifolded configurations with $\langle A_v\rangle$ dominating
over $\langle L\rangle$. See text.
 
Caption to Fig.5

Plot of $\cal L$ at selected values of $q$ against $r$. The logarithmic
scale for $r$ produces the negative of force $-f=\log [r]$.
$N=40$, $q =$ 0.01, 0.1, 0.5, 0.9, 1, 100, $10^4$ (thin lines in the sequence from
left to right). Change of $q$ shifts without change in shape (see text). 
One asymptotic $f_2(q,r)$, eq.(2.27), is shown with a thick line.
Diamonds and crosses: force from eq.(2.33) for boxed
interface with $N=40$ and $N=60$; triangles for $N=40$ with a different
interpolation of $(1/g)(dg/dL)_N$. Also see text.
 
\vfill\eject
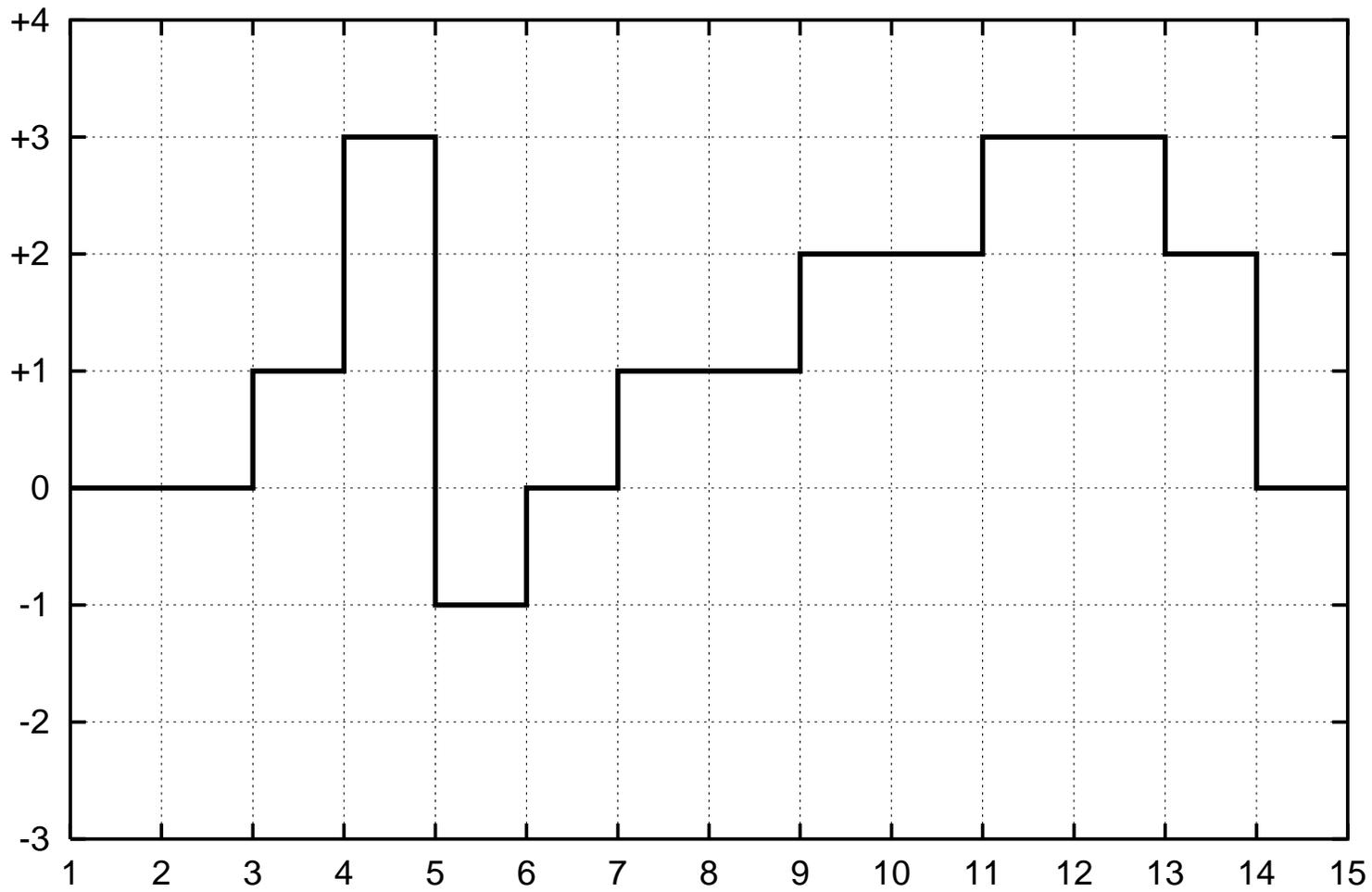
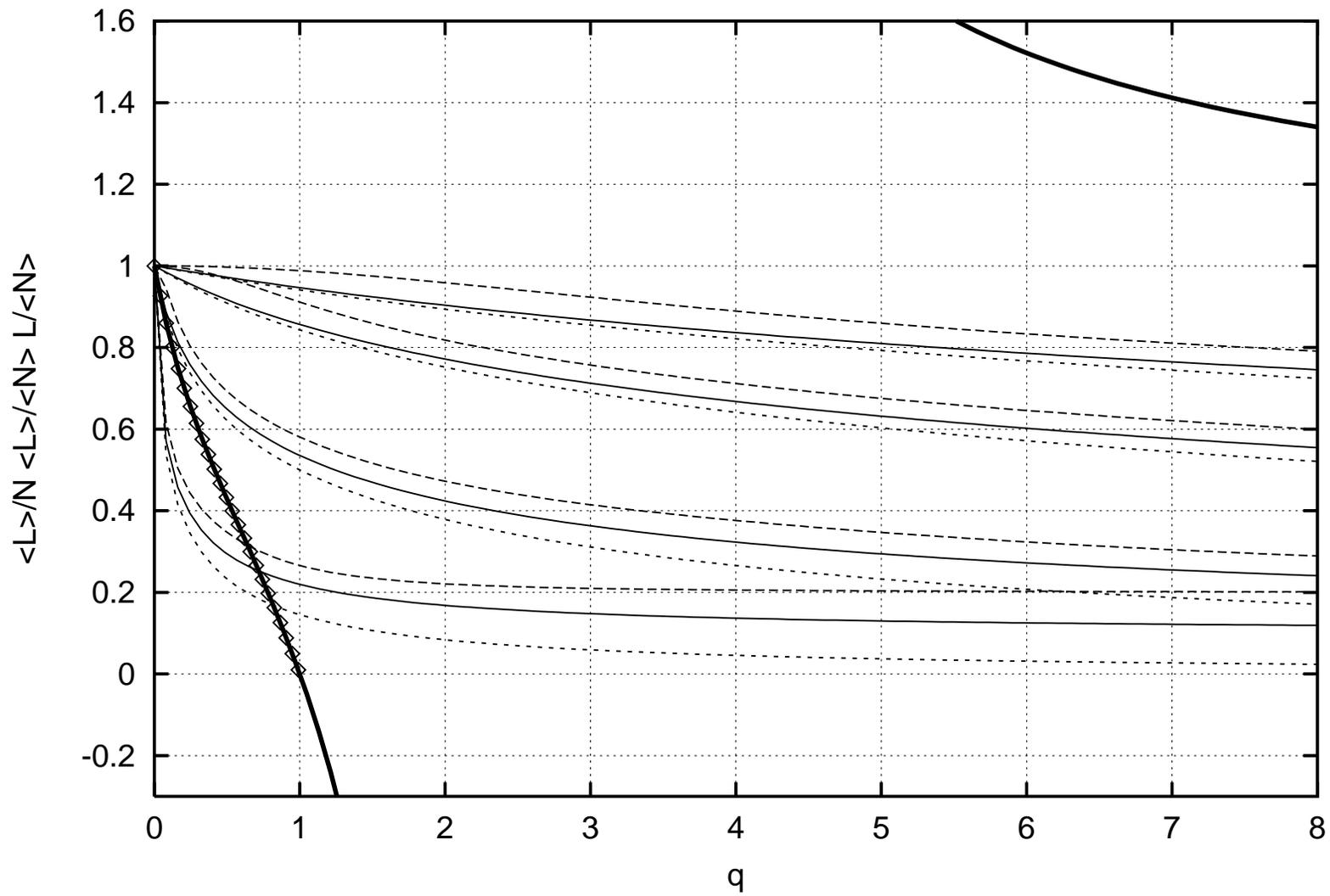
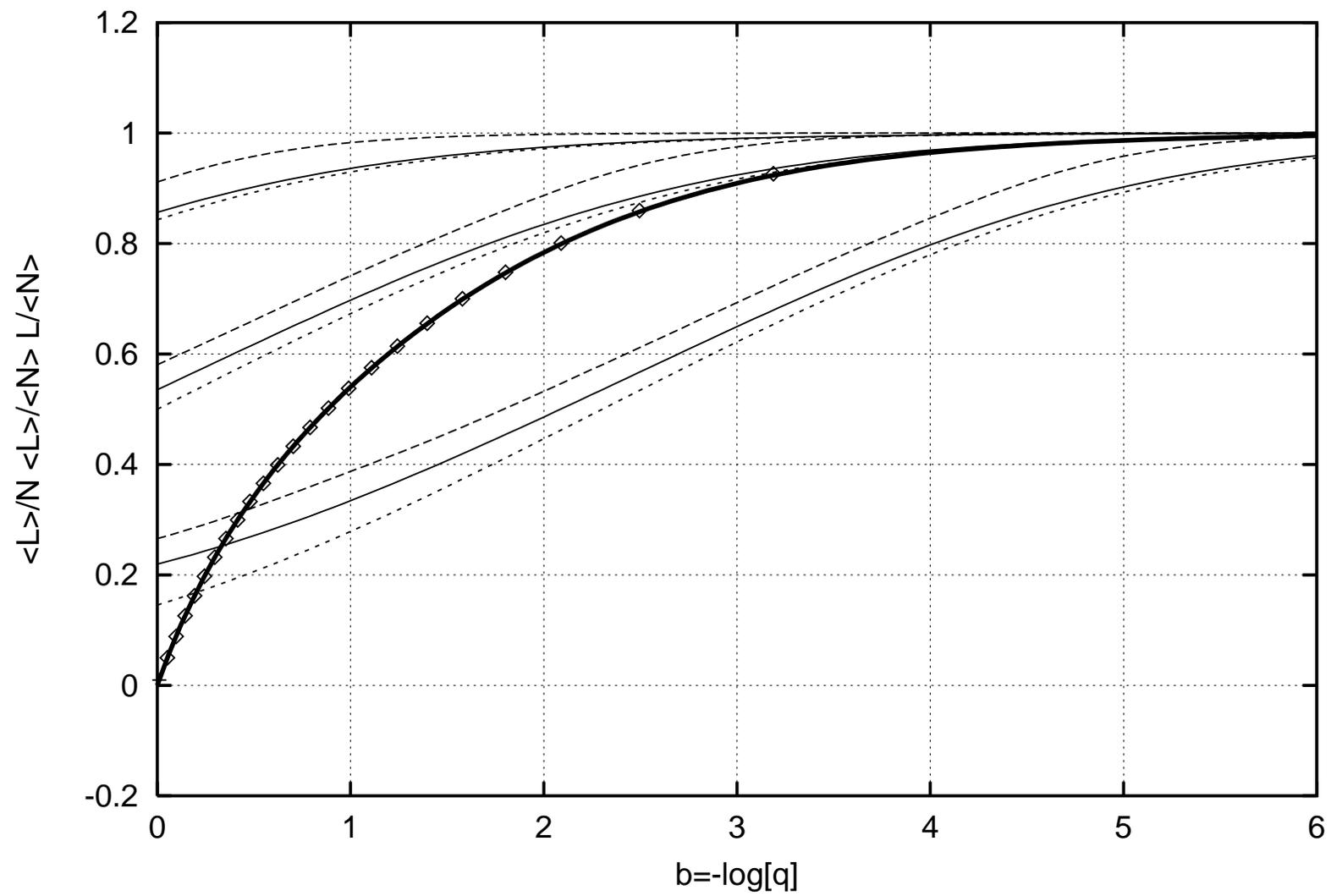
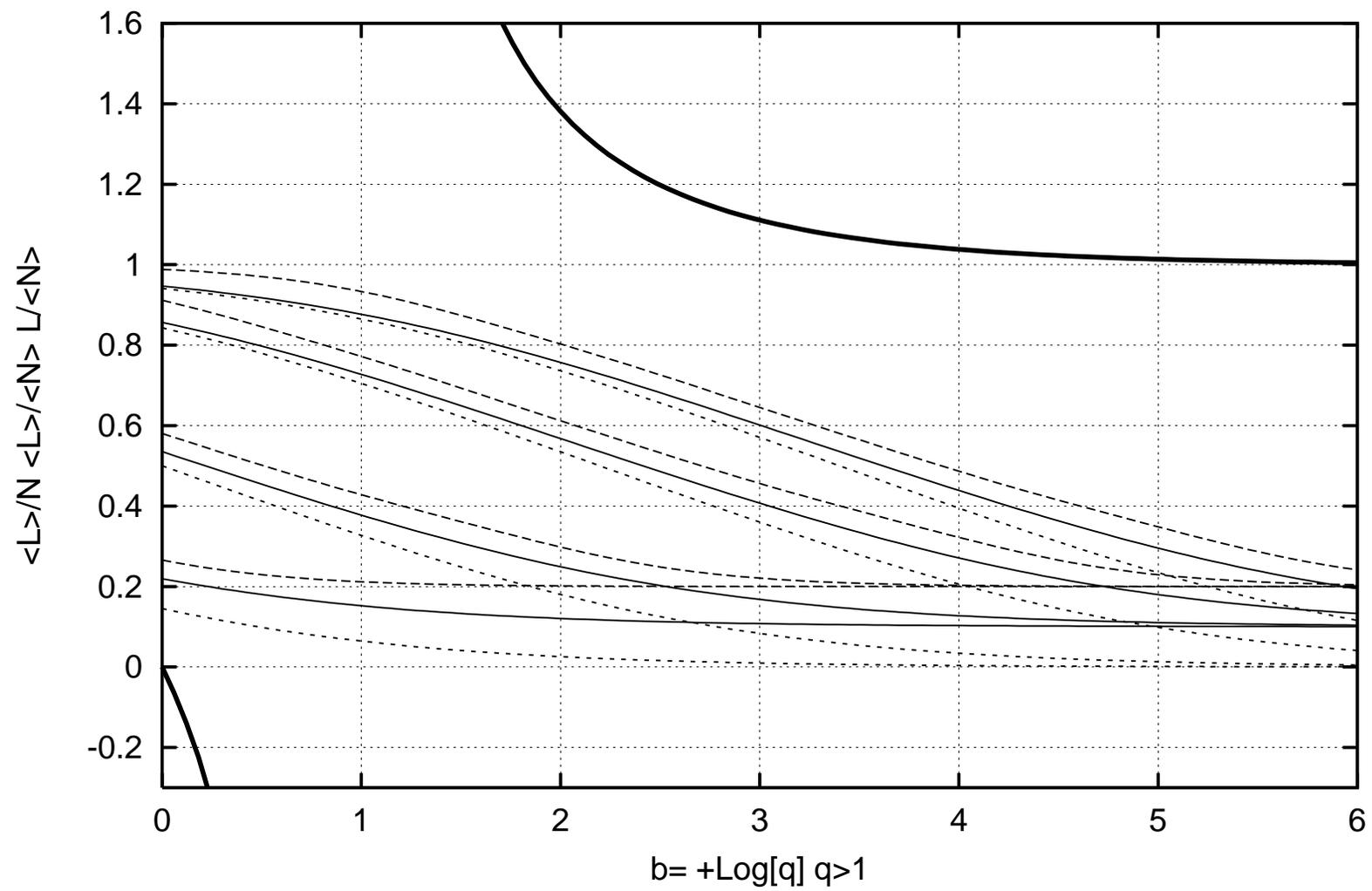
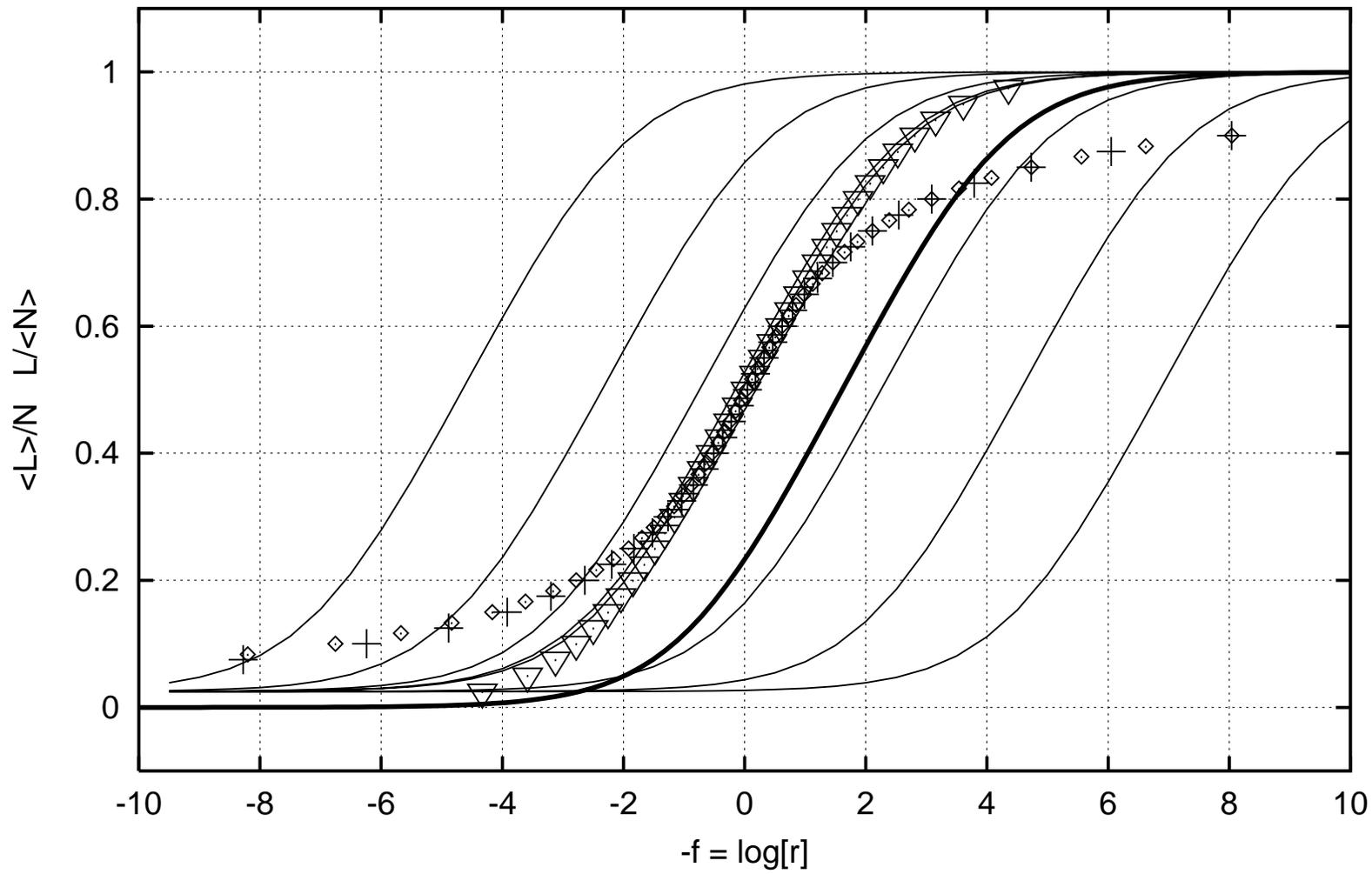
\end